\documentclass[10pt]{article}

\begin{document}

\noindent
\textbf{\Large Continuous Groups with Antilinear Operations}

\bigskip
\noindent
\textbf{J. Koci\'nski$\null^1$ and M. Wierzbicki$\null^2$}

\medskip
\noindent
{\small $\null^1$ Grzybowska 5, m. 401\\
	      00-132 Warszawa, Poland\\
              \texttt{kocinsk@if.pw.edu.pl} }\\

\noindent	   
{\small  $\null^2$  Warsaw University of Technology \\
              Faculty of Physics \\
	      Koszykowa 75 \\
	      00-662 Warszawa, Poland\\
	      \texttt{wierzba@if.pw.edu.pl} \\
}

\begin{abstract}
\noindent
Continuous groups of the form: $G+a_0G$ are defined, where $G$ 
denotes a Lie group and $a_0$ denotes an antilinear operation which fullfils the condition 
$a^2_0=\pm 1$. The matrix algebras connected with the groups $G+a_0G$ are defined.  
The structural constants of these algebras fulfill the 
conditions following from the Jacobi identities.
\end{abstract}


Lie Groups and Algebras

\section{Introduction}

An attempt at an extension of Wigner's investigations \cite{Wigner} of continuous groups
with antilinear operations will be presented.
For a certain type of continuous groups with antilinear operations,
matrix algebras with the commutator product will be defined.
This will be done for continuous groups $G+a_0G$, in which $G$ is a Lie group,
which need not be unitary, and the antilinear element $a_0$ fulfills the condition $a_0^2=\pm 1$.
The matrix algebras connected with the thus specified groups $G+a_0G$ can be defined
in a way which is analogous to that for Lie groups.
The parametrization of the groups $G+a_0G$, where $G$ is a Lie group, requires carefull attention. We will accept the following solution of this problem:
When a Lie group $G$ depends on $n$ essential parameters, the coset $a_0G$ depends on $n+1$ essential parameters --- the $n$ parameters of the Lie group
and an additional parameter which in general is required for completing the matrix algebra.
The required basis element of that algebra is connected with the antilinear element $a_0$.

\section {Continuous groups with antilinear operations}
\label{sec:3}

Wigner \cite{Wigner} considered continuous groups with antilinear operations
$G+a_0G$, where the group $G$ is unitary and $a_0^2\in G$. In the following definition the group $G$ will be a Lie group which need not be unitary, and $a_0^2=\pm 1$.

\noindent
{\bf Definition 2.1.} {\em A type of continuous groups with antilinear operations has the form
$G+a_0G$, where $G$ is a Lie group and $a_0$ is an 
antilinear operation which fullfils the condition $a^2_0=\pm {\bf 1}$}.

\noindent
Before formulating the conditions which any group $G+a_0G$ defined above has to fulfill, we firstly will discuss the problem of the parametrization of the coirreps of that group, and next the problem of the matrix algebra with the commutator product, connected with the above defined group.

\noindent
{\bf Observation 2.1.} From Eqs. (2.11) and (2.12) in \cite{Kocinski6} it is seen that
the matrices $D(g)$ and $D(a)$ of the corepresentation $D\Gamma$,
depend on $n$ essential parameters $\alpha_1,...,\alpha_n$ of the subgroup $G$.
However, for completing the basis of the algebra 
connected with the group $G+a_0G$, with the commutator product,
the basis element connected with the antilinear element $a_0$ can be
required. That basis element can be determined when 
the corepresentation matrices of the coset $a_0G$ depend on
an additional parameter. Such a parameter can be introduced owing to the form
of the equivalence condition of two
corepresentations in Eqs. (2.17) and (2.18) in \cite{Kocinski6}.
The matrices $D'(a)$ then depend on the $n$ essential parameters of the group $G$
and on the parameter $\alpha_0$. 
Consequently, the coirrep $D\Gamma$ depends on $n$ independent parameters
$\alpha_1,...,\alpha_n$ of the linear Lie group $G$, and on the additional
parameter $\alpha_0$, which appears only in the matrices of the coset $a_0G$. 
We will apply the transformation $S$ in Eq. (2.17) from \cite{Kocinski6} which does not alter
the matrices $D(g)$ of the original corepresentation.

It seems that the necessity of introducing the additional parameter is connected
with the existence of two convergence points in the groups $G+a_0G$. In Lie groups, when the essential parameters approach zero values, the representation matrices converge to the unit matrix. In $G+a_0G$ groups there are two points of convergence: the unit matrix $E$ for the Lie group $G$ matrices, and the matrix $D(a_0)$ for the matrices of the coset $a_0G$. In other words, while the local properties of a Lie group $G$ are connected with a small vicinity of the identity transformation, the local properties of the group $G+a_0G$ are connected with the vicinities of two transformations: the identity transformation and the transformation connected with the antilinear element $a_0$.

Depending on the particular group $G+a_0G$, the basis element $X^{\prime}_0$connected with the matrix ${\rm e}^{i\alpha_0}\,D(a_0)$,
either commutes or it does not 
commute with the remaining basis elements of the algebra. 
The question can therefore be posed whether it is legitimate to retain the parameter
$\alpha_0$, with which $X^{\prime}_0$ is connected,
when the algebra is complete without $X^{\prime}_0$. 
Two answers of this question can be considered: (1) The presence of an additional parameter 
in the coset $a_0G$ corepresentation matrices is implied 
by the necessity of completing the algebra. When it turns out that the algebra is complete
without the basis element $X^{\prime}_0$, there is no need for an additional parameter.
There is then no basis element of the algebra which is
connected with the antilinear operation $a_0$.
(2) Any corepresentation can be transformed to the form with the matrices of the coset $a_0G$
depending on the additional parameter. This parameter leads to the basis element $X^{\prime}_0$,
of the matrix algebra which is connected with the antilinear operation $a_0$.

If we accepted the first answer, semi-simple algebras for
the Lie group $G$ could turn into semi-simple algebras for $G+a_0G$. The acceptance of the
second answer implies that 
when $X^{\prime}_0$ commutes with all the remaining basis elements, 
semi-simple algebras of the Lie group $G$ do not turn into semi-simple algebras of the group
$G+a_0G$.  In the following we will accept $\alpha_0$ as 
an essential parameter of the coirrep connected with the group $G+a_0G$, also in the case
when $X^{\prime}_0$ is not required for completing the basis of
the matrix algebra. This means that we accept the second standpoint.

Considering Observation 2.1 we introduce the following four conditions which 
have to be fulfilled by a continuous group with antilinear operations, in the sense of Definition 2.1.

\noindent
(I) The group $G+a_0G$ must have at least one faithful finite-dimensional irreducible
corepresentation $D\Gamma$ of type (a) or (b), of dimension $(n+1)$, i. e. with 
$(n+1)$ essential parameters $\alpha_0,\alpha_1,...,\alpha_n$.

Let the dimension of the irrep matrices $\Delta$ of the subgroup $G$ be $d$. The dimension
of the coirrep $D\Gamma$ is $d$ or $2d$, for $a-$type or $b-$type 
coirreps, respectively.

We define the distance function denoted by $d_1(g,g')$ between 
the elements $g$ and $g'$ of $G$, and the distance function $d_2(a,a')$ between 
the elements $a$ and $a'$
of the coset $a_0G$, 

\begin{eqnarray}
d_1(g,g')=+\sqrt{\sum\limits_{j=1}^{m} \sum\limits_{k=1}^{m}\Big\vert D(g)_{jk}-
D(g')_{jk}\Big\vert^2}
\nonumber\\
d_2(a,a')=+\sqrt{\sum\limits_{j=1}^{m}\sum\limits_{k=1}^{m}\Big\vert D(a)_{jk}-
D(a')_{jk}\Big\vert^2}
\label{eq3:1}
\end{eqnarray}

\noindent
where $m=d$, for $a-$type coirreps, and $m=2d$, for $b-$type coirreps. The parameter $\alpha_0$
does not appear in the distance function $d_2(a,a')$.
These distance functions fulfill the following five conditions: 

\begin{eqnarray}
(1)\,\, d_1(g,g')=d_1(g',g),\quad d_2(a,a')=d_2(a',a)
\nonumber\\
(2)\,\, d_1(g,g)=0,\quad d_2(a,a)=0
\nonumber\\
(3)\,\, d_1(g,g')>0,\,\,\mbox{if}\,\,g\neq g' \quad\mbox{and}\quad d_2(a,a)'>0
\quad\mbox{if}\,\,a\neq a'
\nonumber\\
(4)\,\,d_1(g,g'')\leq d_1(g,g')+d_1(g',g'')\quad\mbox{and}\quad d_2(a,a'')\leq d_2(a,a')+
d_2(a',a'')
\nonumber\\
\mbox{for any three elements of}\,\, G\,\,\mbox{or of}\,\,a_0G
\label{eq3:2}
\end{eqnarray}

\noindent
The two sets of elements $g$ of $G$ and $a$ of $a_0G$ which fulfill the conditions

\begin{equation}
d_1(g,{\bf 1})<\delta_1,\quad\mbox{and}\quad d_2(a,a_0)<\delta_2
\label{eq3:3}
\end{equation}

\noindent
respectively, where $\delta_1$ and $\delta_2$ are real positive numbers, are said to be within 
the sphere of radius $\delta_1$ centered on the unit element ${\bf 1}$, and to be within
the sphere
of radius $\delta_2$ centered on the element $a_0$, respectively. We are dealing with two small
neighbourhoods of ${\bf 1}$ and of $a_0$, respectively. The parameters $\alpha_1,...,\alpha_n$,
on which depend the matrices $D(g)$, representing the Lie subgroup $G$,
are assigned to an $n-$ dimensional Euclidean space ${\cal R}^n$, and
the parameters $\alpha_0,\alpha_1,...,\alpha_n$, on which depend the matrices 
${\rm e}^{i\alpha_0}D(a)$,
representing the coset $a_0G$, are assigned to an $(n+1)-$dimensional
Euclidean space ${\cal R}^{(n+1)}$.

\noindent
(II)$\,$ We fix a $\delta_1>0$ in ${\cal R}^{n}$, and we consider elements
$g$ of $G$ lying in the sphere
of radius $\delta_1$ centered on the unit element ${\bf 1}$. 
At the same time we fix a $\delta_2>0$ 
in ${\cal R}^{(n+1)}$, and consider elements $a$ of $a_0G$, lying in the sphere of
radius $\delta_2$ which is centered on the element $a_0$. The elements $g\in G$ within 
the sphere of radius $\delta_1$ are uniquely parametrized by $n$ real 
parameters $\alpha_1,...,\alpha_n$, and the elements $a\in a_0G$ are uniquely
parametrized by the parameters $\alpha_0,\alpha_1,...,\alpha_n$, when $\alpha_0$ and
$\alpha_0+2\pi p,\, p=\pm 1,2,...,$ are identified.
The matrix $E$, representing the unit element ${\bf 1}$, and
the matrix ${\rm e}^{i\alpha_0}D(a_0)$, representing
the antilinear element $a_0$ are connected with $\alpha_1=...=\alpha_n=0$. 

\noindent
(III)$\,$ There has to exist such $\epsilon_1>0$, that to every point in ${\cal R}^n$
for which
\begin{equation}
\sum\limits_{j=1}^n \alpha^2_j\,<\epsilon^2_1
\label{eq3:4}
\end{equation}

\noindent
there corresponds some element $g$, and there has to exist such $\epsilon_2>0$,
that to every point in ${\cal R}^{(n+1)}$ for which
\begin{equation}
\sum\limits_{j=1}^n \alpha^2_j\,<\epsilon^2_2,\quad {\rm with\,\,a\,\,fixed\,\,\alpha_0}
\label{eq3:5}
\end{equation}

\noindent
there corresponds some element $a=a_0g$. There is a one-to-one correspondence
between elements $g$ of $G$, and points in ${\cal R}^n$, as well as between elements $a$ of
$a_0G$ and points in ${\cal R}^{(n+1)}$, (provided that $\alpha_0$ is identified with
$\alpha_0+2\pi p, p=1,2,...$), satisfying the respective condition in Eq. 
(\ref{eq3:4}) or (\ref{eq3:5}).

\noindent
(IV) $\,$ Each of the matrix elements of the coirrep
$D\Gamma(\alpha_0,\alpha_1,...,\alpha_n)$ must be an analytic function of the parameters
$(\alpha_1,...,\alpha_n)$ for the subgroup $G$, and of these parameters together with
the parameter $\alpha_0$ for the coset $a_0G$. These parameters have to satisfy the respective 
conditions in Eqs. (\ref{eq3:4}) and (\ref{eq3:5}) 
This means that for the subgroup $G$, each of the matrix elements $D_{jk}$
can be expressed as a power series in
$\alpha_1-\alpha^0_1,...,\alpha_n-\alpha^0_n$, for all $(
\alpha^0_1,...,\alpha^0_n)$ fulfilling the condition in Eq. (\ref{eq3:4}), and for the coset
$a_0G$, each of the matrix elements ${\rm exp}(i\alpha_0)D_{jk}$ can be expressed as
a power series in
$\alpha_0-\alpha^0_0, \alpha_1-\alpha^0_1,...,\alpha_n-\alpha^0_n$, for all $\alpha^0_0,
\alpha^0_1,...,\alpha^0_n$ fulfilling the condition in Eq. (\ref{eq3:5}).

Consequently, all the derivatives: $\partial D_{jk}/\partial\alpha_p,\,\,\partial^2 D_{jk}/
\partial\alpha_p\partial\alpha_q,...,$ for the subgroup, and\\
$\partial {\rm exp}(i\alpha_0)D_{jk}/\partial\alpha_p,\,
\partial^2 {\rm exp}(i\alpha_0)D_{jk}/\partial\alpha_p\partial\alpha_q,...,$
for the coset,
have to exist at all points which fulfill Eqs. (\ref{eq3:4}),
and (\ref{eq3:5}), including the points $\alpha_1=...=\alpha_n=0$, and
$\alpha_0=\alpha_1=...=\alpha_n=0$ for the subgroup and for the coset, respectively.
For $a-$type coirreps, we have
$p,q=1,...,n$ and $j,k=1,...,d,$ for the subgroup, and $p,q=0,1,...,n,\,\,$
for the coset, and for $b-$type coirreps we have $p,q=1,...,n;\,\,j,k=1,...,2d,$ 
for the subgroup, and $p,q=0,
1,...,n;\,\,j,k=1,...,2d\,\,$, for the coset. 

\noindent
{\bf Observation 2.2.} The definitions of this Section are valid for the 
"unprimed" form of the corepresentation matrices in Eqs. (2.11) and (2.12) from \cite{Kocinski6}, 
with the factor ${\rm exp}(i\alpha_0)$ in front of the $D(a_0g)$ matrices,
as well as for the "primed" form of the corepresentation matrices 
obtained with the help of the transformations in Eqs. (2.42) and (2.52) from \cite{Kocinski6}.
The "prime" label of the corepresentation matrices $D$ has therefore been 
omitted in points I, II and IV and it will be omitted further on in this Section.

\noindent
{\bf Definition 2.2}. {\em The connected component of the group 
$G+a_0G$, is the maximal set of elements $g$ or $a$ which can be obtained
from each other by continuously varying one or more of the respective matrix elements
$D(g)_{jk}$ or ${\rm e}^{i\alpha_0}D(a)_{jk}$ of the faithful finite-dimensional coirrep $D\Gamma$.}

For both types 
of coirreps, $a$ and $b$, there holds the equivalence condition in Eq. (2.20) from \cite{Kocinski6},
from which we obtain the equality:

\begin{equation}
N\Delta^{\ast}(g^{\prime})=\Delta (g)N
\label{eq3:5a}
\end{equation}

\noindent
where $g^{\prime}\in G$, and for $a_0=K$ we have $g^{\prime}=g^{\ast}$, and for $a_0=\Theta$,
$g^{\prime}=\Theta^{-1}g\Theta$.

For $a-$type coireps, when $a_0=K$, from Eq. (\ref{eq3:5a}) we obtain:
$N\Delta (g)=\Delta (g)N$,
and hence from Schur's lemma we can put $N=E$, with $E$ denoting the unit matrix.
The matrices $D^{\prime}(g)$, $D^{\prime}(Kg)$ and $D^{\prime}(gK)$ reduce to single blocks,
which are $\Delta (g)$, ${\rm exp}(i\alpha_0)\Delta^{\ast}(g)$, and 
${\rm exp}(i\alpha_0)\Delta (g)$, respectively. These can be transformed into one another by
a continuous variation of one or more of the essential parameters. 
If the Lie group $G$ is connected, the group $G+a_0G$ also is connected.
When $a_0\neq K$ and $N\neq E$, we obtain two 
different matrices representing group elements: $\Delta (g)$ and ${\rm exp}(i\alpha)
\Delta (g)N$. We cannot obtain $\Delta (g)$ from $\Delta (g)N$, by a continuous variation of
one or more of the essential parameters.
When the essential parameters approach zero value, the above two matrices converge to
$E$ and to $N\neq E$, respectively. The group $G+a_0G$ then is not connected.

For $b-$type coirreps, the matrices of the coset $a_0G$ due to their form 
cannot be transformed into the matrices of the subgroup $G$
by a continuous variation of one or more of the essential parameters. When the essential
parameters $\alpha_0, \alpha_1,...,\alpha_n$ approach zero values,
the matrices $D(g)$ converge to the unit matrix $E$, and the matrices
${\rm e}^{i\alpha_0}D(a_0g)$ and ${\rm e}^{i\alpha_0}D(ga_0)$ to the matrix $D(a_0)$ 
in Eq. (2.49) from \cite{Kocinski6}. According to Definition 2.2, for $b-$type coirreps
the groups $G+a_0G$ are not connected.

\noindent
{\bf Definition 2.3}. {\em For $a-$type coirreps we define the d-dimensional matrices $X_1,...
X_n$, connected with the subgroup $G$, and the $d-$dimensional matrices
$X^{\prime}_{n+1}$, $\ldots$, $X^{\prime}_{2n}$ and $X^{\prime}_0$, connected with the coset $a_0G$,
by their elements}

\begin{equation}
(X_p)_{jk}=\Big(\frac{\partial D(g)_{jk}}{\partial\alpha_p}\Big)_{\alpha_1=...=\alpha_n=0}
\,;\quad p=1,...,n
\label{eq3:6}
\end{equation}
\begin{equation}
(X^{\prime}_q)_{jk}=\Big(\frac{\partial {\rm e}^{i\alpha_0} D(a_0g)_{jk}}{\partial\alpha_q}
\Big)_{\alpha_0=\alpha_1=...=\alpha_n=0}\,;\quad q=0,1,...,n
\label{eq3:7}
\end{equation}

\noindent
{\em where $D(g)_{jk}$ and ${\rm e}^{i\alpha_0}D(a_0g)_{jk},\,j,k=1,...,d$ denote 
the elements of the respective coirrep matrices}.

\noindent
{\bf Corollary 2.1.} {\em For a-type coirreps, when all the matrices 
$X^{\prime}_q,\,q=1,...,n,$ are linearly dependent on the matrices
$X_p,\,p=1,...,n,$, the $(n+1)$ matrices $X^{\prime}_0, X_1,...,X_n$} span 
an $(n+1)-$dimensional vector basis. 

\noindent
{\em Proof}. The proof is analogous to that for Lie groups in \cite{Cornwell}.

\noindent
{\bf Definition 2.4}. {\em For $b-$type coirreps, 
the $2d-$dimensional matrices $X_1$, $\ldots$, $X_n$,
$X^{\prime}_{n+1}$, $\ldots$, $X^{\prime}_{2n}$, and $X^{\prime}_0$ are defined by their elements}

\begin{equation}
(X_p)_{jk}=\Big(\frac{\partial D(g)_{jk}}{\partial\alpha_p}\Big)_{\alpha_1=\alpha_2=...=
\alpha_n=0}; \quad p=1,...,n
\label{eq3:8}
\end{equation}
\begin{equation}
(X^{\prime}_q)_{jk}=\Big(\frac{\partial {\rm e}^{i\alpha_0}D(a_0g)_{jk}}{\partial\alpha_q}
\Big)_{\alpha_0=\alpha_1=...=\alpha_n=0};\quad q=0,1,...,n
\label{eq3:9}
\end{equation}

\noindent
{\em where ${\rm e}^{i\alpha_0}D(a_0g)_{jk},\,j,k=1,2,...,2d$, denote the elements of
the coirrep matrices of the coset $a_0G$, and where the matrices $D(a_0g)$ do not depend
on the parameter $\alpha_0$}.

\noindent
{\bf Corollary 2.2.} {\em For b-type coirreps, the matrices $X_1,...,X_n,X^{\prime}_{n+1},...,
X^{\prime}_{2n}$, and $X^{\prime}_0$ defined by Eqs. 
(\ref{eq3:8}) and (\ref{eq3:9}) span a $(2n+1)-$dimensional real vector space.}

\noindent
{\em Proof}. Because of their form, the matrices
$X^{\prime}_{(n+1)},...,X^{\prime}_{(2n)}$, $X^{\prime}_0$,
connected with the elements $a$ in $a_0G$, 
always are linearly independent of the matrices $X_1,...,X_n$, connected with
the subgroup $G$. We know that the matrices $X_1,...,X_n$,  are linearly independent
\cite{Cornwell}. It suffices to demonstrate the linear independence of the $(n+1)$ matrices
$X^{\prime}_0, X^{\prime}_{n+1},..., X^{\prime}_{2n}$, connected with the coset $a_0G$.
We have to show that the only solution of the equation
\begin{equation}
\Big(\sum\limits_{j=n+1}^{2n} \lambda_j X^{\prime}_j\Big)+\lambda_0X^{\prime}_0=0,
\quad\mbox{with all $\lambda'$s real}
\label{eq3:10}
\end{equation}

\noindent
is $\lambda_{n+1}=\lambda_{n+2}=...=\lambda_{2n}=\lambda_0=0$. The respective proof is
analogous to that in \cite{Cornwell}, for a Lie group $G$.

\noindent
{\bf Conjecture 2.1}. {\em In a complex algebra, the matrices $X_1,...,X_n,X^{\prime}_0$, 
span an $(n+1)-$dimensional vector space.}

\noindent
{\bf Definition 2.5}.
{\em For the matrices $X_1,...,X_n,X^{\prime}_{n+1},...,X^{\prime}_{2n}$, 
and $X^{\prime}_0$, we define the commutator products $[A,B]=AB-BA$.}

For the Lie group $G$ we have:

\begin{equation}
[X_p,\,X_q]=\sum\limits_{r=1}^n {\overline c}^{\,r}_{pq}\,X_r,\quad p,q,r=1,...,n
\label{eq3:11}
\end{equation}

\noindent
where ${\overline c}^{\,r}_{pq}$ are the structural constants. For the remaining
commutator products we introduce

\noindent
\textbf{Definition 2.6}
\textit{The commutator of two basis vectors connected with the coset $a_0G$
is equal to a linear combination of basis vectors connected with the subgroup~$G$}

\begin{equation}
[X^{\prime}_p,\,X^{\prime}_q]=\sum\limits_{r=1}^n {\overline d}^{\,r}_{pq}\,X_r,
\quad p,q=0,n+1,...,2n; \quad r=1,...,n
\label{eq3:12}
\end{equation}

\noindent
\textit{and the commutator of a basis vector $X_p$, connected with the subgroup $G$,
with a basis vector $X^{\prime}_q$, connected with the coset $a_0G$, is equal to a linear 
combination of basis vectors connected with that coset,}

\begin{equation}
[X_p,X^{\prime}_q]=\sum\limits_r {\overline e}^{\,r}_{pq}X^{\prime}_r,
\quad p=1,...,n;\quad q=0,n+1,...,2n; \quad r=0,n+1,...,2n
\label{eq3:13}
\end{equation}

\noindent
\textit{where the structural constants
${\overline d}^{\,r}_{pq}$ and ${\overline e}^{\,r}_{pq}$ 
are antisymmetric with respect to the interchange of the
indices $p$ and $q$.}

The definitions in Eqs. (\ref{eq3:11}), (\ref{eq3:12}) and (\ref{eq3:13}) establish
a correspondence between the results of products of elements in the group $G+a_0G$, and the results of the respective commutator products of the basis elements of the matrix algebra.

\noindent
{\bf Corollary 2.3.} 
{\em 
From the Jacobi identity for the double commutator $[[X_p,X_q],X_r]$,
in Lie algebras we obtain the known relation between the structural constants
${\overline c}^{\,r}_{pq}$ in Eq. (\ref{eq3:11}).
From the three Jacobi identities connected with the double commutators: 
$[[X_p,X_q],X^{\prime}_r],\, [[X_p,X^{\prime}_q],X^{\prime}_r]$ and 
$[[X^{\prime}_p,X^{\prime}_q],X^{\prime}_r]$, we obtain on the basis of
Eqs. (\ref{eq3:11}),
(\ref{eq3:12}) and (\ref{eq3:13}) the respective three relations between the structural
constants ${\overline c}^{\,s}_{pq}$, ${\overline d}^{\,s}_{pq}$ and ${\overline e}^{\,s}_{pq}$}:
\begin{eqnarray}
{\overline c}^{\,s}_{pq}{\overline e}^{\,t}_{sr}-
{\overline e}^{\,s}_{qr}{\overline e}^{\,t}_{ps}+
{\overline e}^{\,s}_{pr}{\overline e}^{\,t}_{qs}=0
\nonumber\\
{\overline e}^{\,s}_{pq}{\overline d}^{\,t}_{sr}+
{\overline d}^{\,s}_{qr}{\overline c}^{\,t}_{sp}-
{\overline e}^{\,s}_{pr}{\overline d}^{\,t}_{sq}=0
\nonumber\\
{\overline d}^{\,s}_{pq}{\overline e}^{\,t}_{sr}+
{\overline d}^{\,s}_{qr}{\overline e}^{\,t}_{sp}+
{\overline d}^{\,s}_{rp}{\overline e}^{\,t}_{sq}=0
\label{eq3:15}
\end{eqnarray} 

\section{Conclusions}

Wigner \cite{Wigner} considered the continuous groups $G+a_0G$ with a unitary group $G$ and the antilinear element $a_0$, with $a_0^2\in G$.
The matrix algebras with commutator product connected with those groups were not considered. We have considered the continuous groups $G+a_0G$, where $G$
is a Lie group, which need not be unitary, and $a_0$ is an antilinear operation
which fulfills the condition $a_0^2=\pm 1$.
The $a$-type and $b$-type irreducible corepresentations of the groups $G+a_0G$ were employed for the determination of
the respective matrix algebras with commutator product. Some of the general properties of those algebras were determined.

We have presented the corepresentation theory without the assumption of the unitarity of
the subgroup $G$ of the group $G+a_0G$, where $a_0$ denotes an antilinear operation.
This was done for coirreps
of $a-$type or $b-$type (types $1$ or $2$, respectively, in \cite
{Wigner}). To the matrices representing
the elements of the coset $a_0G$, an additional parameter $\alpha_0$ is assigned
by means of the equivalence transformation of two coirreps.
It then is possible to define the basis element
$X^{\prime}_0$, connected with the matrix ${\rm e}^{i\alpha_0}D(a_0)$. This basis
element in general is required for completing the algebra connected
with the group $G+a_0G$. There are cases, however, depending on the Lie group $G$,
and on the type of the antilinear element $a_0$, 
when $X^{\prime}_0$ commutes with the remaining basis elements
of the respective algebra. It then is not indispensible for completing that algebra.
The parameter $\alpha_0$ is included into the set of essential parameters 
of the group $G+a_0G$ in any case. Consequently,
all the matrices ${\rm e}^{i\alpha_0}D(a_0g)$, with $g\in G$, belong to an $(n+1)-$dimensional 
parameter space, while the matrices $D(g)$ belong to an $n-$dimensional parameter space of 
the Lie group $G$.

There appears a characteristic difference between the properties of Lie groups $G$, and of
groups $G+a_0G$. In Lie groups, when the essential parameters approach zero
values, the representing matrices converge to the unit matrix $E$. 
In $G+a_0G$ groups, there are two points of convergence:
the unit matrix $E$ for the Lie group $G$ matrices, and the matrix $D(a_0)$
for the matrices of the coset $a_0G$. In other words,
while the local properties of a Lie group $G$ are connected with a small vicinity of 
the identity transformation, the local properties of the group $G+a_0G$ 
are connected with the vicinities of two transformations: the identity transformation and
the transformation connected with the antilinear element~$a_0$.

\section*{Acknowledgments}
We are very much indebted to Professor Zbigniew Oziewicz from the Universidad
Nacional Aut\'onoma de M\'exico for the discussions concerning the mappings with antilinear operations, and for a critical reading of the paper.


\begin{thebibliography}{}

\bibitem{Birman}
Birman, J. L. (1984). {\em Theory of Crystal Space Groups and Lattice Dynamics},
Springer-Verlag, Berlin.

\bibitem{Bir}
Bir, G. L. and G. E. Pikus (1972). {\em Symmetry and deformation effects in semi-conductors},
(in Russian), Nauka, Moscow.

\bibitem{Bradley}
Bradley, C. J., and A. P. Cracknell (1972). {\em The Mathematical Theory of Symmetry in Solids. 
Representation Theory for Point Groups and Space Groups.} Clarendon Press, Oxford.
\bibitem{Cornwell}
Cornwell, J. F. (1984). {\em Group Theory in Physics}, Vol. I, II, Academic Press, New York.

\bibitem{Dimmock}
Dimmock, J. O., and R. G. Wheeler. (1964). In {\em The Mathematics of Physics and Chemistry},
Vol. {\bf 2}, Eds. H. Margenau and G. M. Murphy, Van Nostrand, New York.

\bibitem{Fleming}
Fleming, W. (1977). {\em Functions of Several Variables, Second Edition}, Springer-Verlag,
New York, Heidelberg and Berlin.

\bibitem{Flugge}
Fl\"ugge, S. (1964). {\em Lehrbuch der Theoretischen Physik, Vol. IV: Quantentheorie I}, 
Springer, Berlin.

\bibitem{Gross}
Gross, F. (1993). {\em Relativistic Quantum Mechanics and Field Theory}, Wiley,
New York.

\bibitem{Hammermesh}
Hammermesh, M. (1962). {\em Group Theory and its Application to Physical Problems},
Addison-Wesley Publishing Company, Inc., Reading, Massachusetts. 

\bibitem{Jar1}
Jaroszewicz, A., P. Koci\'nski, G. T{\c e}cza and J. Koci\'nski. (1989). The spin structure of 
neodymium in a magnetic field, {\em Physica}, {\bf B 156-157}, 756-758.

\bibitem{Kocinski1}
Koci\'nski, J. (1983). {\em Theory of Symmetry Changes at Continuous Phase Transitions},
Elsevier Science Publishers, Amsterdam.

\bibitem{Kocinski2}
Koci\'nski, J. and K. Osuch. (1987). Symmetry Changes at the Tricritical Point in
Metamagnets, {\em Phase Transitions}, {\bf 10}, 151-180.

\bibitem{Kocinski3}
Koci\'nski, J. (1990). {\em Commensurate and Incommensurate Phase Transitions}, 
Elsevier Science Publishers, Amsterdam.

\bibitem{Kocinski4}
Koci\'nski, J. (1995). Wigner's Theorem for Non-Unitary Symmetry Groups. In {\em Symmetry
and Structural Properties of Condensed Matter}, 77-88; Eds. T. Lulek, W. Florek and S. Wa\char 32 lcerz,
World Scientific, Singapore.

\bibitem{Kocinski5}
Koci\'nski, J. (1999). Corepresentations in Lattice Vibrations Theory. In {\em Symmetry and 
Structural Properties of Condensed Matter}, 435-452; Eds. T. Lulek, B. Lulek and A. Wal. World
Scientific, Singapore.

\bibitem{Kocinski6}
Koci\'nski, J. and Wierzbicki, M. (2009). The Corepresentations of Continuous Groups, arXiv: 0905.4828v1 [math-ph]




\bibitem{KocinskiP}
Koci\'nski, P. (1992). The Eigenfunctions of the Pauli Hamiltonian for an iron crystal in the 
ferromagnetic phase, {\em J. Phys. Chem. Solids}, {\bf 55}, 1189-1195.

\bibitem{Kovalev1}
Kovalev, O. V. (1983). Peculiarities in applications of corepresentation theory to the problem
of lattice vibrations, (in Russian). {\em Preprint deposited in VINITI}, No. 1089, Kharkov.

\bibitem{Kovalev2}
Kovalev, O. V. (1983). The method of induced corepresentations and the invariant expression
for energy in the problem of lattice vibrations, (in Russian). {\em Preprint deposited in
VINITI}, No. 1090, Kharkov.

\bibitem{Kovalev3}
Kovalev, O. V. (1983). Closed expressions for macroscopic parameters in the method of 
induced corepresentations in lattice dynamics, (in Russian). {\em Preprint deposited in VINITI},
No. 1091, Kharkov, and {\em Summary} in {\em Low Temperature Physics}, {\bf 9}, 10.

\bibitem{Kovalev4}
Kovalev, O. V. (1983). The chain-of-secular-equations method in crystal lattice dynamics,
(in Russian). {\em Preprint deposited in VINITI}, No. 1092, Kharkov, and {\em Summary} in
{\em Low Temperature Physics}, {\bf 9}, 10.

\bibitem{Kovalev5}
Kovalev, O. V. (1985). Real forms of small representations in phase transition theory.
{\em Physics of Metals}, (in Russian), {\bf 59}, 5, 1032-1033.

\bibitem{Kovalev6}
Kovalev O. V., and A. G. Gorbanyuk. (1985). {\em The Irreducible Corepresentations of 
Magnetic Space Groups with Anti-rotation}, (in Russian), Naukova Dumka, Kiev.

\bibitem{Kovalev7}
Kovalev, O. V. (1986). {\em Irreducible and Induced Representations and Corepresentations 
of Fedorov Groups}, (in Russian), Nauka, Moscow.

\bibitem{Kuratowski}
Kuratowski, C. (1948). {\em Topologie I, Deuxieme \'Edition Revue et Augment\'ee},
Warszawa.
 
\bibitem{Naimark}
Naimark, M. A. (1975). {\em Theory of Group Representations}, (in Russian), Nauka,
Moscow.

\bibitem{Pontryagin}
Pontryagin, L. (1966). {\em Topological Groups: Second Edition}, Gordon and Breach.



\bibitem{Streitwolf}
Streitwolf, H. W. (1967). {\em Gruppentheorie in der Festk\"orperphysik}, Academische
Verlagsgesellschaft, Leipzig.

\bibitem{Wigner}
Wigner, E. P. (1959). {\em Group Theory and its Application to the Quantum Mechanics
of Atomic Spectra}, Academic Press, New York.

\end{thebibliography}
\end{document}